\documentclass[11pt, twocolumn]{article}

\usepackage[left=0.7in, right=0.5in, top=1in, bottom=1in]{geometry}
\usepackage[colorlinks]{hyperref}
\usepackage{amsmath}
\usepackage{amssymb}
\usepackage{garamondlibre}
\usepackage[backend=biber, citestyle=numeric-comp, sorting=none]{biblatex}
\usepackage{orcidlink}
\usepackage{graphicx}
\usepackage[font=small]{caption}

\definecolor{KUred}{RGB}{144, 26, 30}
\definecolor{randomblue}{HTML}{4b7aaf}

\hypersetup{citecolor=randomblue, linkcolor=KUred, urlcolor=black}

\DeclareCaptionLabelSeparator{colon}{\textcolor{KUred}{\textbf{{:\space}}}}
\DeclareCaptionLabelFormat{redbold}{\textcolor{KUred}{\textbf{#1 #2}}}
\begin{document}

\twocolumn[
  \begin{@twocolumnfalse}
    
    \title{\vspace{-40pt} A repeating sequence of simple rules creates a universal computer}

    \author{
    Riz Fernando Noronha \orcidlink{0009-0007-2923-3835}
    }

    \date{\textit{Niels Bohr Institute, University of Copenhagen, Copenhagen 2200, Denmark} \\ \vspace{10pt}
    Corresponding author: \href{mailto:noronha@nbi.ku.dk}{noronha@nbi.ku.dk}\vspace{-10pt}}

    \maketitle

    \renewcommand{\abstractname}{\large Abstract\\}

    \begin{abstract}
    \normalsize
    We show that the Turing complete rule 110 in elementary cellular automata can be emulated by alternating three simple symmetric rules in space. As the constituent rules themselves are trivial, this shows that diversity alone can generate emergent phenomena. The rules cannot distinguish left from right, and the chirality of rule 110 is created by their arrangement. Random initial conditions lock on to rule 110 over 90\% of the time, explained through the motion of domain walls between desynchronized regions. Adding random noise creates errors that seed domain walls, whose fraction scales with the square root of noise strength, explained through a simple rate equation. This points to a route to complexity through repeated, perturbed units, common in biology, and shows that in cellular automaton models of quantum field theory, broken symmetries such as parity need not be present in the local rules.
    \end{abstract}
    
    \vspace{20pt}
  \end{@twocolumnfalse}
]

\section{Introduction}

Elementary cellular automata (ECA), proposed by Stephen Wolfram \cite{wolfram1983Statistical, wolfram1984Universality, wolfram2002NKS}, are considered to be the simplest systems which can give rise to complex, emergent behavior from identical local rules. Perhaps the most famous of the 256 rules is rule 110, which can support gliders \cite{martinez2006Gliders} and is Turing complete \cite{cook2004Universality}, even efficiently so \cite{nearyWoods2006Pcomplete}. Here, complexity is a property of a single rule copied onto every site.

Life often builds complexity differently, from repeated, near-identical units. In the fruit fly, each body segment is patterned by a few genes expressed in stripes of cells in a fixed order, repeating along the embryo. Signalling between neighbouring cells keeps this pattern stable, corrects errors in it, and gives each segment a front and a back \cite{vonDassow2000Polarity, albertOthmer2003Boolean}. At the molecular scale, gene duplications often split a function between them, so each copy is necessary \cite{force1999DDC}. In V-ATPase, this turned a ring of two protein types into a ring of three, in a fixed arrangement \cite{finnigan2012VATPase}.

\begin{figure}[h!]
    \centering
    \includegraphics[width=\linewidth]{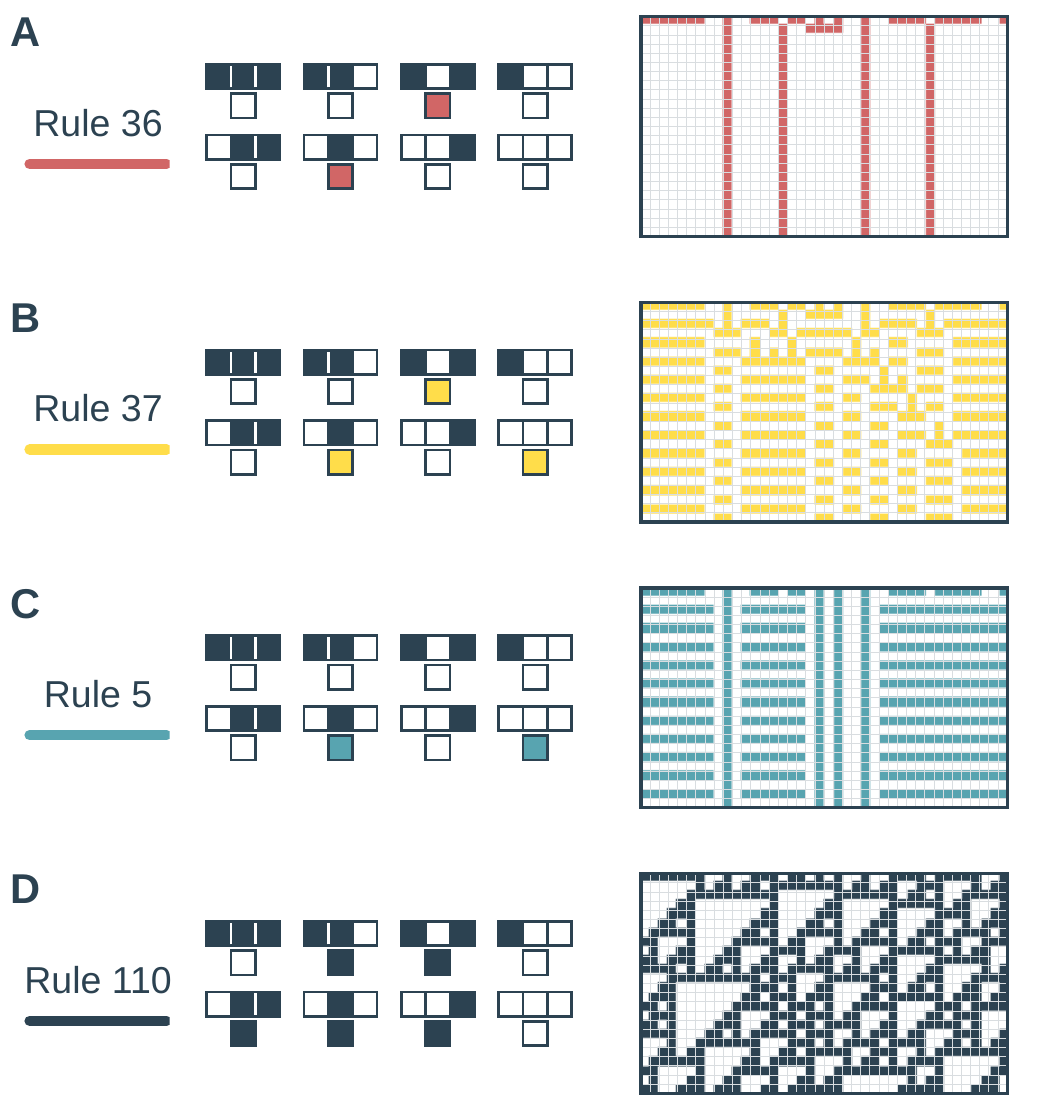}
    \caption{Rule tables. Rules 36, 37, and 5 (A-C) are all mirror symmetric, and result in trivial oscillating or stationary patterns. Rule 110 (D) displays remarkable complex structure.}
    \label{fig:ruleTables}
\end{figure}

Non-uniform automata, in contrast to ECA, have different cells follow different rules \cite{cattaneo2009NonUniform, dennunzio2012NonUniform, bhattacharjee2020Survey}. They are known to be able to compute where uniform ones in the same space cannot \cite{sipper1995Quasi, sipper1996Coevolving, sipper1997Evolution, sipper1998Nonuniformity}: for instance, alternating two rules temporally can solve the density classification problem which no single two-state rule can \cite{fuks1997Density, landBelew1995Density}.

Riedel and Zenil wrote rule 110 as a composition in time of simpler rules \cite{riedelZenil2018Primality}. However, one of their components, rule 118, is left-right asymmetric, thus explicitly showing the symmetry-breaking expected in rule 110.

By Curie's principle, an asymmetry in an effect must already be present among its causes \cite{curie1894Symmetrie}. In physics, the weak interaction violates parity \cite{leeYang1956Parity, wu1957Parity}, and 't Hooft has proposed that quantum mechanics emerges from a deterministic cellular automaton, noting that it is hard to foresee how the symmetries of nature appear in its rules \cite{thooft2016CAI}.

In this letter, we show that rule 110 can be emulated by three mirror-symmetric, individually trivial rules alternating in space, robust to noise, where the chirality of rule 110 is then created by their arrangement.

\section{Emulating rule 110}

Unlike rule 110, which is asymmetric and needs to be defined by its entire rule table, we build it from the three rules that can be described more simply below:
\begin{enumerate}\setlength{\itemsep}{0em}
    \item Rule 36: become active only if both neighbours are the opposite color of the cell.
    \item Rule 37: the same as rule 36, but also activate if all sites are inactive.
    \item Rule 5: Become active only if both outer neighbours are inactive.
\end{enumerate}
All three are mirror-symmetric, and eventually result in trivial frozen or oscillatory patterns (\autoref{fig:ruleTables}).

\begin{figure}[h!]
    \centering
    \includegraphics[width=\linewidth]{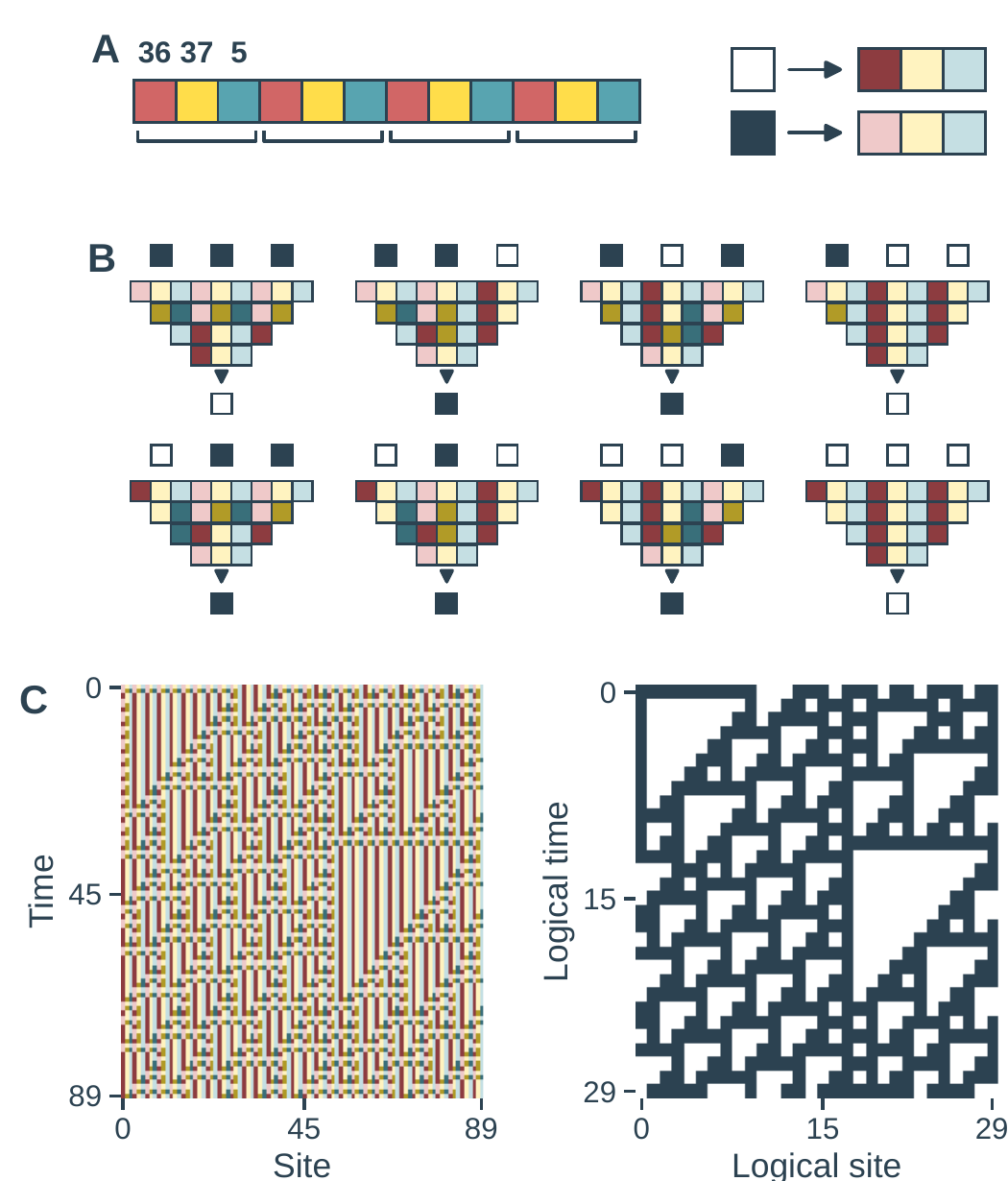}
    \caption{Combining the three simple rules can give rise to rule 110. A: The tiling pattern of the rules, and the assignment of [100] to [0] and [000] to [1]; B: The effective rule 110 table in the combined system, showing how each input ultimately reduces to the output expected of rule 110; C: A simulation trace, which upon rescaling, recreates rule 110.}
    \label{fig:translatingRules}
\end{figure}

We assign rule 36 to sites where $i\bmod 3=0$, rule 37 where $i\bmod 3=1$ and rule 5 to $i\bmod 3 = 2$ \footnote{Rule 36 can be replaced by rule 54, as the two rules only differ in positions that are never visited.}. If we then encode logical [0] as the block [100] and logical [1] as [000], three steps of the mixture equal one step of rule 110 (\autoref{fig:translatingRules}).

Because the emulation is exact, the mixture inherits the universality of rule 110: any computation in Cook's construction \cite{cook2004Universality} can be run on the mixture by writing each cell as a three-cell block, at a constant cost of three in both space and time, so the efficiency shown in \cite{nearyWoods2006Pcomplete} is also preserved.

The construction necessarily requires a minimum of three rules at period three. This is due to symmetry: rule 110 distinguishes left from right, though the component rules do not. A pattern of two symmetric rules repeating ($a, b, a, b,\dots$) can also not make this distinction \footnote{Neither can ($a, a, b, \dots$), as reversing it gives a shifted copy. The shortest arrangement with two rules needs a period of at least 6, ($a, a, b, a, b, b,\dots$)}, and thus the shortest arrangement needed involves three rules ($a, b, c, a,\dots$). The chirality of rule 110 thus is created by the arrangement, instead of being supplied as in previous work \cite{riedelZenil2018Primality}.

\section{Random initial conditions}

\begin{figure}[h!]
    \centering
    \includegraphics[width=\linewidth]{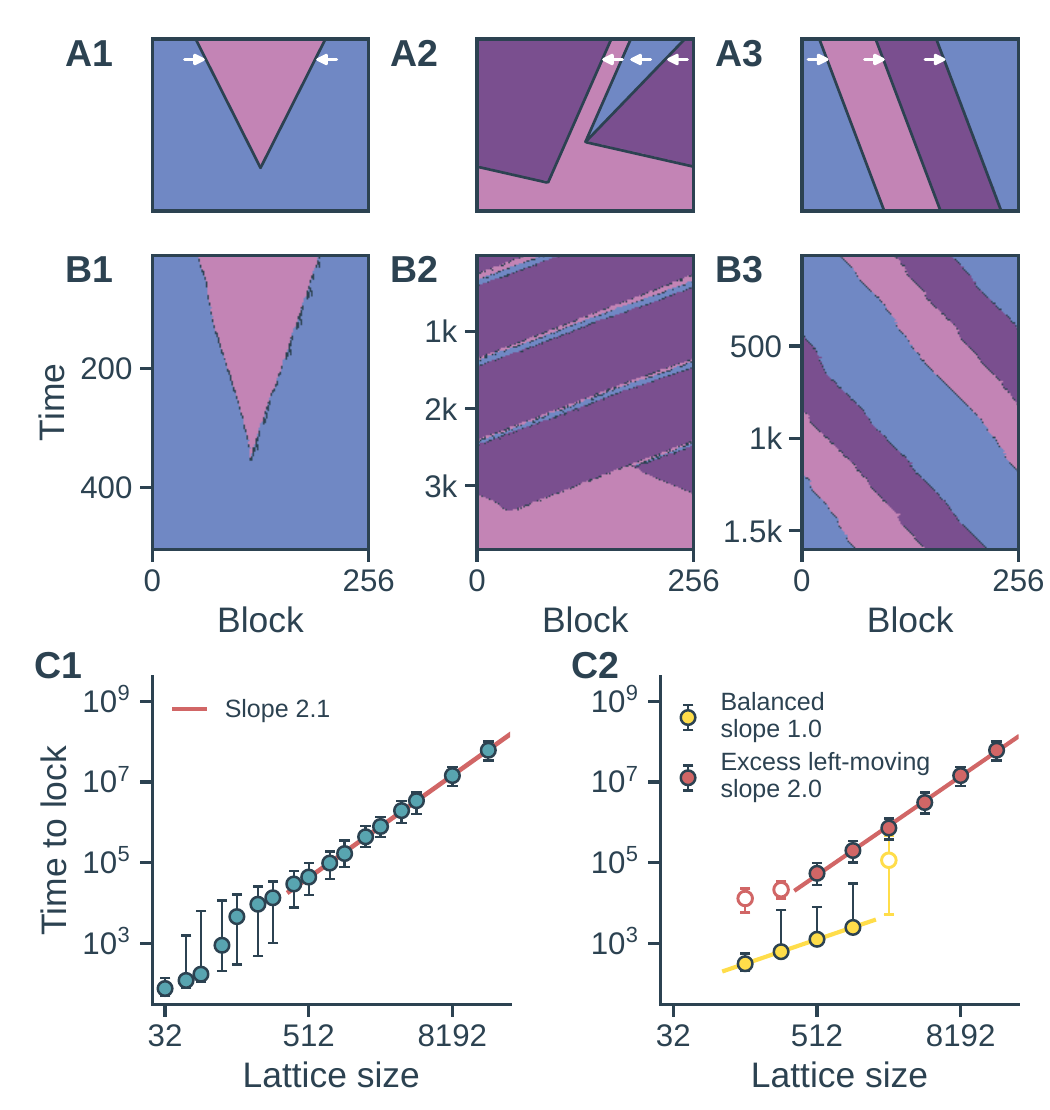}
    \caption{Schematic (A) and simulations (B) of the system's time evolution for different scenarios based on the domain walls. Different colors represent regions of space emulating rule 110 in different phases. If domains are balanced (left and right), they converge and annihilate (A1/B1). If domains move leftwards, one is pinned, and eventually takes over (A2/B2). If domains are moving rightwards, they never approach each other, and the system never locks (A3/B3). C: Locking times versus system size, for the systems that do lock. C1: All systems, showing an approximate power law of $N^{2}$, C2: Segregated by mechanisms: balanced scales with $N$, while left-moving walls (which grow in abundance for high $N$) scale with $N^2$.}
    \label{fig:randomICs}
\end{figure}

While coded initial conditions replicate rule 110, we examine the system's robustness to random initial conditions, by performing simulations on lattices with periodic boundary conditions. We observe that the mixture largely self-organizes into rule 110. For lattices of length $N\geq 256$, $\approx 91-100\%$ of runs lock into rule 110, with the locking time scaling approximately as a power law $\sim N^{2}$ (\autoref{fig:randomICs}). Around $0-9\%$ of runs, depending on lattice size, never locked to rule 110 and were trapped \footnote{It is possible that some of the `trapped' runs simply had long transients, and ran out of time, especially at 16,384 blocks.}.

This trapping is due to the formation of domains. The construction replicates rule 110 every third step, but the lattice is divided into regions that locally replicate rule 110, but out of phase with one another (steps 0, 3, 6, etc., as opposed to 1, 4, 7, etc.). Domain walls are observed to drift towards domains that are a single beat behind them. Thus, depending on the initial conditions, one can get domain walls that either move leftwards or rightwards \footnote{This description of the dynamics in terms of domains and the walls between them follows the computational-mechanics picture of cellular automata \cite{hansonCrutchfield1992Attractor, crutchfieldHanson1993Turbulent}, but in contrast to the diffusive movements of rule 18, the walls here move ballistically.}.

If the domain walls are balanced (one of each kind), then left-moving and right-moving walls will collide and annihilate, leaving the system locked to a homogeneous phase to replicate rule 110. However, due to the chirality, domain walls moving leftwards behave quite differently from those moving to the right. Walls moving leftwards pin a small domain against another wall, then eliminate another domain, allowing the pinned domain to expand to the right and eventually take over the system. On the other hand, walls moving rightward that are spaced apart never approach each other, and continue moving without ever getting pinned.

\section{Noisy dynamics}

In order to analyze robustness in the dynamics, we give each bit a chance to flip with probability $\varepsilon$ each step. We can then count the number of domain walls, unbroken stretches of lattice blocks that do not follow the code that emulates rule 110. The domain wall density is observed to scale with the square root of noise.

A simple argument can explain the square root behavior. When a bit is flipped, it has a probability $(1-p)$ to be healed after a few iterations, or a probability $p$ to create two domain walls, one on each side \footnote{In a way, this is simpler than the random initial conditions, as there it is possible to create more domain walls of one kind.}. Each block has 3 bits, and so walls are created with rate $6 p \varepsilon$. The domain walls can drift with velocity $v$, half to the left, and half to the right. A wall drifting left approaches another drifting right at relative speed $2v$, and when they collide, both walls are eliminated. Thus, we can write a simple rate equation for the density of the domain walls, $n$:

\begin{equation}
    \frac{\mathrm d n}{\mathrm d t} =  6 p \varepsilon - 4 v \left(\frac{n}2 \right)^2
\end{equation}

Walls are created randomly at a constant rate, and die upon collision. Solving this gives

\begin{equation}
    n^\ast = \sqrt{\frac{6 p \varepsilon}{v}}
\end{equation}

\begin{figure}[h!]
    \centering
    \includegraphics[width=\linewidth]{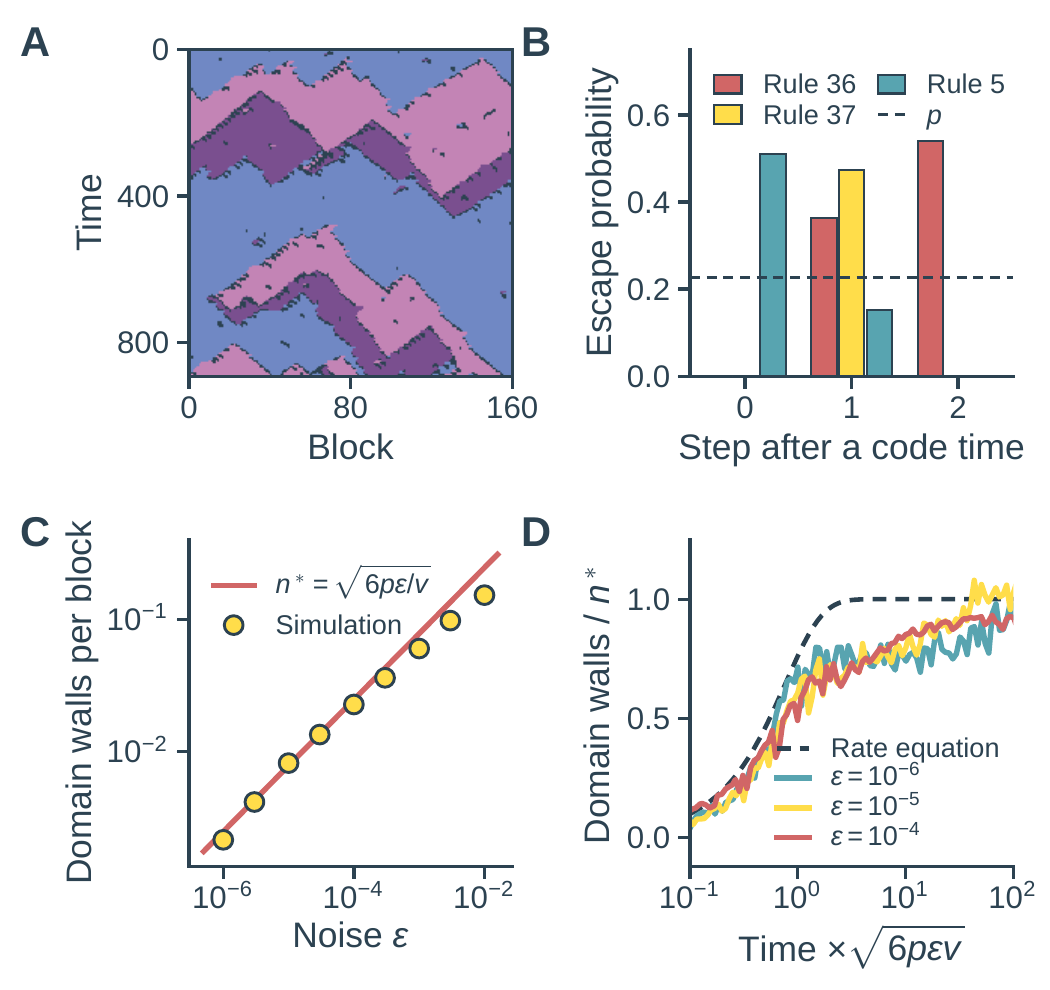}
    \caption{Behavior of the system under noise. A: A plot of the errors propagating in the model with random bit flips. Domain walls form in pairs, and spread with a constant drift. B: The probability for an error to lead to the formation of a domain wall, for each rule. Four out of nine never split into walls. $p$ is estimated as the average. C, D: The steady-state fraction of domain walls as a function of noise, showing $\sqrt\varepsilon$ scaling (C) and the rescaled fraction of domain walls over time (D), with the rate equation prediction using parameters for $p=0.23$ and $v=(v_L + v_R)/2=0.2275$ estimated from simulations.}
    \label{fig:noise}
\end{figure}

This explains the square root behavior. We measured the parameters from the simulation to be $p\approx 0.23$ (averaged over the three phases of the code). The walls do indeed move apart, but at different velocities, $v_R\approx0.219$ and $v_L\approx0.236$ blocks per step. The dynamics of domain walls, together with the rate equation predictions, are visualized in \autoref{fig:noise}. For large noise, walls lie in the vicinity of each other and interfere, and thus fail to act independently.

The final approach to $n^\ast$ is slower than this rate equation implies, possibly because walls that survive the early collisions mostly move in the same direction, and thus meet only rarely.

\section{Discussion}

Rule 110 is a standard example of emergence: using a simple rule, and many different lattice sites that follow the same rule, a beautiful pattern that is Turing complete can be generated. However, there appears to be a `threshold' of the simplicity of the rule, as mirror symmetric rules cannot generate such complex phenomena. In our construction, the missing ingredient is supplied by the arrangement, and the chirality is explicitly seen in the speeds of domain walls. As complex physics mainly uses techniques from statistical physics which disproportionately deal with ensembles of identically interacting agents, it often glosses over emergence arising from heterogeneity, as discussed here. Many interacting agents with heterogeneous but simpler rules can be a separate route to emergence, similar to how heterogeneity enhances collective behaviour in other systems \cite{tessone2006Diversity, sanchezpuig2023Heterogeneity, tlustyLibchaber2025Cascade}.

One might hope that diversity saves information, with several simple rules replacing one complex rule. Unfortunately, that is not the case here. Describing a rule as its family plus its member within the family \cite{rissanen1978Modeling, grunwald2007MDL}, the mixture of simple rules was observed to use more information than rule 110 alone. Diversity thus redistributes rule information among the agents rather than reducing it. It is possible that in larger rule spaces, where uniform rule tables become more expensive, this is not the case and emergence from diversity is `simpler' in terms of information complexity.

Our results also have implications for biology. Rule 37 activates on the conditions that activate rules 36 and 5 combined (101, 010, 000). Thus, they can be read as copies of rule 37 which each lost a different entry in the rule table. This is the route by which duplicated genes split an ancestral function \cite{force1999DDC}, and by which the V-ATPase ring gained its third subunit \cite{finnigan2012VATPase}. Constructive neutral evolution similarly holds that complexity often arises because it is easily reached \cite{stoltzfus1999CNE, gray2010Irremediable}, consistent with the mixture using more information than rule 110.

In 't Hooft's cellular automaton interpretation, splitting the lattice into even and odd sublattices is treated as a convenient tool for model building \cite{thooft2016CAI}. Our results show that sublattices carry symmetry, as two cannot break parity, but three can. The parity violation of the Standard Model thus need not be written into the local rules of such an automaton, but could arise from their arrangement, though our rules are irreversible (unlike most considered there).

The construction may also be of practical use. An elementary rule that depends only on its own state and the number of active neighbors is mirror symmetric. Such rules show up in many physical systems, such as evenly spaced Rydberg atoms \cite{wintermantel2020RydbergQCA} or engineered cells signalling by diffusion \cite{tamsir2011NOR, regot2011Distributed}, where a chiral rule is hard to engineer directly, but can be built from them. The noise analysis gives a rough error budget: $L$ blocks stay free of domain walls when $n^\ast L < 1$, i.e. $\varepsilon < v/(6pL^2)\approx 0.16/L^2$.

Our work complements Sipper's finding that non-uniformity reduces the neighborhood size a task needs \cite{sipper1996Coevolving, sipper1998Nonuniformity}; neighborhood size is a different resource from description length. It also complements with Israeli and Goldenfeld's observation that coarse-grained rules can be simpler than microscopic ones \cite{israeliGoldenfeld2004Irreducibility, israeliGoldenfeld2006Coarse}.

One could also consider our construction as a larger cellular automaton. By grouping the three constituent cells into a single `super-cell', the construction can be viewed as a uniform eight-state cellular automaton \cite{mazoyerRapaport1999Grouping, delorme2011BulkingI}. The usefulness of the construction, however, lies in the triviality and symmetry of its components.

\section{Conclusion}

In conclusion, we show that a repeating sequence of three simple, mirror-symmetric rules can emulate rule 110, creating a universal computer whose chirality is supplied by the arrangement of its rules. The mixture usually self-organizes into rule 110 from random initial conditions, with failures explained by the motion of domain walls between desynchronized regions. Under noise, errors seed pairs of domain walls whose density scales as $\sqrt{\varepsilon}$, captured by a simple rate equation. These results show that complexity can emerge robustly from the arrangement of simple parts.

\section{Acknowledgements}

I thank Jakob Schauser, Thomas Heimburg, and Ulrik Hvid for discussions. The code for all analysis and visualization in the manuscript was written by Claude Opus 5.5. Claude Opus 5.5 also helped with the literature search. All AI-generated results were verified by the author. This work is supported by the Novo Nordisk Fonden Grant No. NNF21OC0065542

\printbibliography

\end{document}